# QCD spectroscopy

Don Weingarten[a]

[a]IBM Research, P.O.B 218,
Yorktown Heights, NY 10598, USA

I review recent work on the calculation of hadron masses, decay constants and wave functions.

## 1. INTRODUCTION

Exactly what set of topics is supposed to go into the spectroscopy review talk seems to vary a bit from one lattice conference to the next. My definition of the territory this year is masses, decay constants and wave functions for both the valence (quenched) approximation and full QCD. I will include glueballs but not hadrons containing heavy quarks and will restrict myself mainly to work done during the last year.

The spectroscopy of low lying quark states and glue states I think remains a critical part of the QCD. An unavoidable component of showing that QCD actually does explain the strong interactions is to show that it accounts for the masses of low lying hadrons. QCD has been assigned the homework problem of hadron spectroscopy just as early quantum mechanics was assigned to work on the spectrum of hydrogen and other light atoms. In addition, whatever algorithms can prove themselves in reproducing the known features of low lying spectroscopy can then be applied with greater confidence to extract predictions from QCD for things we do not know yet. The masses of the lightest glueballs, of course, are not known. A reliable lattice calculation of these numbers seems likely to be a required first step toward the experimental identification of glueballs.

The overall picture of the present state of hadron spectroscopy is roughly as follows. For spectroscopy including quark-antiquark vacuum polarization, life is still pretty difficult. Calculations are restricted to fairly small statistical ensembles and comparatively large lattice spacing and quark mass. Significantly more machine power, perhaps a factor of 100 or more, or a real improvement in algorithms, are needed to get reliable estimates for the zero lattice spacing, infinite volume limits of most properties of light hadrons. For full QCD calculations and valence approximation calculations tuned to the same values of $m_\pi/m_\rho$, lattice spacing in physical units and lattice period in physical units, and with reasonably large lattice period, corresponding hadron masses, at the parameter values for which calculations have been done so far, agree within statistics. Full QCD and valence approximation calculations, with parameters tuned to agree at sufficiently large volume, yield masses which disagree if the the lattice volume is then made much smaller. The difference in volume dependence between full QCD and the valence approximation has been shown to be a consequence of the nonzero vacuum expectation value which occurs in full QCD for gauge loops closed around a space-direction lattice period. Except for very small lattices, these loops have zero expectation in the valence approximation.

For spectroscopy in the valence approximation, on the other hand, there are now calculations for Wilson quarks of masses and decay constants extrapolated to physical quark mass, zero lattice spacing and infinite volume. Ratios of several light hadron masses found in these limits differ from experiment by less than 6%, with statistical errors of 8% and less, and are statistically consistent with experiment. Although chiral perturbation theory suggests that the valence approximation will behave anomalously at sufficiently small quark mass, it now appears that this problem would only disrupt the hadron mass ratios which have been calculated if they were evaluated at quark masses below the light quark masses. Valence approximation meson decay constants extrapolated to physical quark mass, zero lattice spacing and infinite volume fall typically about 15% below experiment. This underestimate ranges in significance from about 1 to 3 standard deviations. The valence approximation is expected, however, to be most reliable for quantities determined primarily by the low momentum behavior of the chromoelectric field. Decay constants are proportional to the absolute square of meson wave functions at the origin and are thus sensitive to the high momentum behavior of the chromoelectric field. A simple renormalization group estimate suggests that the error in the valence approximation's treatment of the high momentum behavior of the chromoelectric field will lead to decay constants falling below those of the full theory.

Two independent, infinite volume, continuum limit valence approximation calculations have now also been reported for the scalar glueball mass. The two data sets agree within statistical errors before extrapolation, but the two groups extrapolate differently to zero lattice spacing and get answers which are not quite consistent. The calculation with higher statistical weight, using about eight times as many configurations, predicts a mass of $1740 \pm 70$ MeV, favoring $f_0(1710)$ as the scalar glueball.

The subjects I am going to cover will be organized as follows:

1. Technical issues

   - Extracting hadron masses from propagators
   - Volume dependence
   - Valence approximation versus full QCD
   - Wave functions, state vectors

2. Calculations with fixed lattice spacing

   - Valence approximation
   - Full QCD

3. Valence approximation zero lattice spacing limits

   - Masses
   - Decay constants
   - Glueball masses

## 2. TECHNICAL ISSUES

### 2.1. Extracting Masses from Propagators

At Lattice 92 the QCDPAX collaboration [1] presented calculations checking for high mass contamination in the values of hadron masses reported by several other groups. The lightest mass contributing to a propagator should be extracted from the propagator's fall off at asymptotically large time separations, and picks up high mass contamination from excited intermediate states if it is obtained at time separations which are too small. The value which QCDPAX found for the rho mass with Wilson quarks in the valence approximation using 200 gauge configurations on a lattice $24^3 \times 54$ at $\beta$ of 6.0 and $k$ of 0.1550 was 3 standard deviations below the value which the APE collaboration found at the same $\beta$ and $k$ using 78 gauge configurations on a lattice $24^3 \times 32$ [2]. Both groups have now run new calculations with larger ensembles [3,4]. The new calculations of the rho mass are both in agreement with the QCDPAX value from Lattice 92. The original disagreement appears to me to have been caused by a statistical fluctuation in the first APE ensemble. The fluctuation caused the appearance of an effective mass plateau at a somewhat smaller time separation and therefore with a somewhat larger mass value than occurs in the new data.

At Lattice 92 Ukawa [5] questioned whether high mass contamination might not be present in some of the masses which I discussed from calculations on GF11 [6]. A comparison of masses found from propagators for several different choices of hadron sink operators, however, provides evidence that our numbers probably do not have significant high mass contamination. This subject will be discussed below in Sect. 4.1.

### 2.2. Volume Dependence

New data, extending earlier results [7], on the volume dependence of hadron masses in full QCD was presented by the MILC collaboration [8].

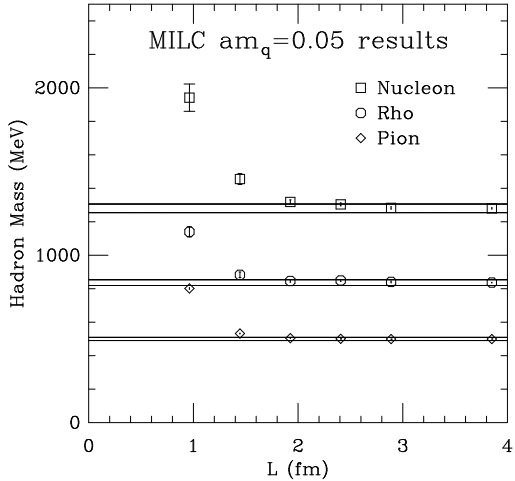

Figure 1. Volume dependence of hadron masses found by the MILC collaboration at $\beta$ of 5.415.

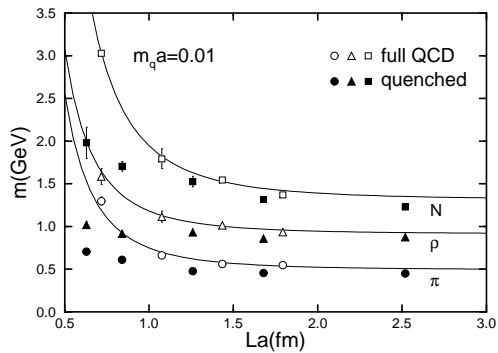

Figure 2. Volume dependence of hadron masses found by the Kyoto-Tsukuba collaboration at $\beta$ of 5.7.

These calculations were done with full QCD for two flavors of Kogut-Susskind fermions. A summary of the run parameters is shown in Table 1. For the range of parameters examined, values of $m_N$ and $m_\rho$ within 2% of their infinite volume limits required space direction lattice periods greater than about 2.5 fm. The volume dependence of hadron masses for $m_q a$ of 0.05 is shown in Figure 1. The horizontal lines indicate intervals within 2% of infinite volume masses. The lattice spacing was chosen by setting the pion mass to 500 MeV, since in this case $m_\pi/m_\rho$ is nearly

| $\beta$ | $m_q a$ | $S$ | $m_\pi/m_\rho$ | $m_\rho a$ |
|---|---|---|---|---|
| 5.415 | 0.050 | $4-16$ | 0.597(1) | 1.023(2) |
| 5.445 | 0.025 | $8-16$ | 0.489(2) | 0.918(4) |
| 5.470 | 0.0125 | $12-16$ | 0.367(3) | 0.883(6) |

Table 1

Parameters of MILC collaboration runs, on lattices $S^3 \times 24$ with $m_\pi/m_\rho$ in each case given for the largest lattice.

| $\beta$ | $m_q a$ | $S$ | $m_\pi/m_\rho$ | $m_\rho a$ |
|---|---|---|---|---|
| | | valence approximation | | |
| 6.0 | 0.02 | $6-24$ | 0.643(4) | 0.520(3) |
| 6.0 | 0.01 | $6-24$ | 0.513(7) | 0.465(6) |
| | | full QCD | | |
| 5.7 | 0.02 | $8-20$ | 0.692(5) | 0.492(3) |
| 5.7 | 0.01 | $8-20$ | 0.586(11) | 0.418(7) |

Table 2

Parameters of runs by the Kyoto-Tsukuba collaboration, on lattices $S^3 \times 24$ with $m_\pi/m_\rho$ and $m_\rho a$ in each case given for the largest lattice.

equal to the observed value of $m_K/m_K^*$.

Data on the volume dependence of valence approximation masses [9] and a comparison with the volume dependence of full QCD masses [10] was reported by the Kyoto-Tsukuba collaboration . The parameters of these runs are shown in Table 2. For the largest values of space direction period, the values of $m_\pi/m_\rho$ and $m_\rho a$ for the valence approximation run at $m_q a$ of 0.01 appear roughly comparable to those of the full QCD run at the same quark mass, and a similar approximate equality holds for the runs at $m_q a$ of 0.02. Data for full QCD and for the valence approximation with $m_q a$ of 0.01 is shown in Figure 2. In the valence approximation, however, $m_\pi$, $m_\rho$ and $m_N$ are found to approach their infinite volume values more rapidly than they do in full QCD. The origin of this difference is shown to be related to the difference between the behavior in full QCD and in the valence approximation of the variables $P_i$, the trace of the product of gauge links closed around the lattice period in space direction $i$. The values of $P_i$ affect the contribution to quark propagators of quark paths wrapping around the lattice period. In the valence approximation, for lattice periods larger than the inverse of the deconfining temperature, the familiar $Z_3$ symmetry of the QCD action without fermions prevents $P_i$ from acquiring a vacuum expectation value. In full QCD this symmetry is broken by the fermion terms in the action, and $P_i$ has a nonzero vacuum expectation value which falls continuously to zero as the lattice period becomes larger. By changing the boundary conditions of the fermions entering the action from periodic to antiperiodic, the sign of the vacuum expectation of $P_i$ can be changed. The Kyoto-Tsukuba collaboration shows that this change causes hadron masses to go from decreasing with volume to increasing with volume. This result, along with similar information obtained varying the boundary conditions on the quarks entering hadron propagators, establishes that the difference between the vacuum expectation of $P_i$ in the valence approximation and in the full theory accounts for the differences in volume dependence.

The volume dependence in hadron masses in full QCD reported by the MILC collaboration and by the Kyoto-Tsukuba collaboration are consistent with each other. For all but the largest values of space direction lattice period, both data sets shown significantly stronger volume dependence than predicted by Lüscher's rigorous asymptotic formula [11]. Both sets of results are consistent with the behavior

$$m(L) = m(\infty) + \frac{c}{L^3} \qquad (1)$$

predicted for intermediate values of $L$ by the model of Ref. [12]. Lüscher's formula embodies the effect of meson exchange between a hadron and its nearest periodic image. The model of Ref. [12] is intended to take into account in addition the finite extension of a hadron's wave function, or equivalently the momentum dependent form factor appearing in a hadron's coupling to exchanged mesons. For lattice periods significantly larger than the extension of a hadron's wave function, the model of Ref. [12] is designed to reproduce Lüscher's formula. Fits of the Kyoto-Tsukuba data for full QCD to Eq. (1) are shown in Figure 2.

## 2.3. Valence Approximation versus Full QCD

The valence approximation may be viewed as replacing the momentum-dependent color dielectric constant arising from quark-antiquark vacuum polarization with its zero-momentum limit [13]. This approximation might be expected to be fairly reliable for low-lying baryon and meson masses determined by the low momentum behavior of the chromoelectric field.

Missing from the valence approximation, however, are couplings of vector mesons to their decay channels. For the spin-3/2 baryon multiplet, these couplings are present but altered from their values in full QCD [14]. A calculation by Leinweber and Cohen [15] attempts to estimate the effect of this omission on $m_\rho$. These authors consider also the effect of obtaining $m_\rho$ by the linear extrapolation down to light quark masses of values of vector masses calculated with heavier quarks. Leinweber and Cohen use a simple local coupling between the rho and two pions, combined with a momentum cutoff in the otherwise divergent pion loop integral entering the rho propagator. The cutoff is intended to model the momentum dependence of the coupling between the rho and pions which, in the real world, would cause the loop integral to converge. The error in the valence approximation value for $m_\rho$ arising from the missing loop combined with extrapolation from the range of quark masses used in Ref. [6], is estimated to be between 0 and $-25$ MeV, thus less than 3%. For a linear extrapolation of the vector meson mass as a function of valence quark mass in full QCD, working from a quark mass interval extending upward from the point corresponding to a 500 MeV pion, the error arising from extrapolation alone is estimated to be $-10$ to $-20$ MeV.

A discussion in Ref. [6] of the extrapolation of hadron masses produces as a byproduct an additional piece of evidence suggesting that the coupling of light hadrons to decay channels has a relatively small effect on their masses in full QCD and therefore gives rise to a relatively small error when omitted or altered in the valence approximation. Take the up and down quark masses to be equal, for simplicity, to a single normal quark mass, $m_n$. Then expand the masses of the hadrons in the vector, spin-1/2 and spin-3/2 baryon multiplets as power series in the mass of the strange quark, $m_s$, around the point at which the strange quark mass and normal quark mass are equal, keeping only the leading linear term. Applying flavor SU(3) symmetry at the point with $m_s$ equal to $m_n$ then gives relations between the masses of hadrons composed of a single flavor of quark and hadrons composed of two different flavors of quarks. For the vector multiplet we obtain, for example,

$$m_\rho[(m_n + m_s)/2, (m_n + m_s)/2] =$$
$$m_{K^*}(m_n, m_s). \quad (2)$$

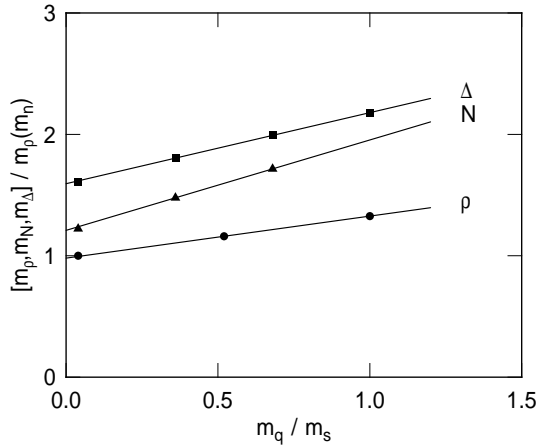

Figure 3. Linear extrapolations to determine nonstrange vector, spin-1/2 and spin-3/2 baryon masses from the masses of their strange partners.

In addition, in the valence approximation, $m_\phi$ is given by $m_\rho(m_s, m_s)$. Corresponding sets of relations hold in the baryon multiplets.

Using these relations, the mass of the member of each multiplet containing no strange quarks can be obtained by linear extrapolation from the masses of the multiplet members containing one or more strange quark. Extrapolating from experimentally observed masses in this way requires only the quark content of each hadron involved and does not depend on the actual values of quark masses. These extrapolations are, essentially, just a reinterpretation of the Gell-Mann-Okubo mass formulas. Figure 3 shows extrapo-lations for the vector mesons, spin-1/2 baryons and spin-3/2 baryons. Leinweber and Cohen's calculation for full QCD may be viewed as an estimate of the error of the extrapolation for the vector multiplet. In place of their error value of $-10$ to $-20$ MeV, the error actually obtained applying this extrapolation to the mass data from experiment is $-4$ MeV, or about 0.5%. For the spin-1/2 baryon multiplet the error is $+13$ MeV, about 1.4%, and for the spin-3/2 baryon multiplet the error is $+10$ MeV, about 0.8%. In two of these multiplets, however, decay widths vary drastically among the strange hadrons to which the linear fits are made and are generally much smaller than the decay width of the nonstrange hadron whose mass is found by the extrapolation. In the vector meson multiplet, the rho width is 149 MeV, while the $K^*$ and phi used to find the rho mass have widths of 49.8 MeV and 4.4 MeV, respectively. The members of the spin-1/2 baryon multiplet are stable with respect to the strong interaction. In the spin-3/2 baryon multiplet, the $\Delta$ has a width of about 115 MeV, while $\Sigma^*$, $\Xi^*$ and $\Omega$ used to find the $\Delta$ mass have widths of 36 MeV, 10 MeV and 0 MeV, respectively. The $\Omega$ does not decay by the strong interaction. It is hard to imagine how the masses of the broad rho and $\Delta$ could be extrapolated so accurately from their strange partners, whose widths are much narrower and vary significantly, if hadron masses were altered by more than a percent or so by their decay widths.

Calculations of the difference between the valence approximation and full QCD in the limit of small quark masses have been reported by Bernard and Golterman [16] and by Sharpe [17]. For hadrons composed of quarks which are sufficiently light, these authors show that chiral perturbation theory suggests the valence approximation will behave pathologically. As a consequence of the absence of vacuum polarization, the $\eta'$ mass goes to zero along with the pion mass as quark masses are made small. Virtual $\eta'$ loops become divergent at low momentum as the quark mass is made small and contribute additional logarithms of the quark mass not present in chiral perturbation theory for full QCD and not multiplied by as many powers of $m_\pi$ as would occur in the full theory. For the behavior of the pion mass at small quark mass, which in full QCD is

$$m_\pi^2 = cm_q, \tag{3}$$

Sharpe has summed a collection of such singular diagrams for the valence approximation and obtained

$$m_\pi^2 = c'm_q^{\frac{1}{1+\delta}}, \tag{4}$$

$$\delta = \frac{\mu^2}{N_f(4\pi f_\pi)^2}. \tag{5}$$

Here $\mu$ is a measure of the amplitude for the quark-antiquark pair in the $\eta'$ to annihilate into gluons. In full QCD $\mu$ becomes the mass of the $\eta'$ in the limit of zero quark mass and would be about 900 MeV. Using this estimate, along with 93 MeV for $f_\pi$, the value of $\delta$ becomes about 0.2.

There is, however, no reason to assume $\mu$ has the same value in full QCD and in the valence approximation. A recent Monte Carlo evaluation of $\mu$ in the valence approximation [18] yielded $0.700 \pm 0.050$ GeV, giving $\delta$ of about 0.12. Fits of Eq. (4) to data to be discussed in Sect. 4.1 for Wilson quarks [6] and for Kogut-Susskind [25] quarks, give values of $\delta$ shown in Table 3. For Wilson quarks, $m_q$ in these fits is $(2k)^{-1} - (2k_c)^{-1}$ where $k_c$ is the critical hopping constant at which $m_\pi$ becomes zero. The value of $k_c$ was taken as one of the adjustable parameters in each fit. The negative values of $\delta$ for two out of three of the fits with Wilson quarks imply that the corresponding $m_q$ are above the range in which the argument leading to Eq. (4) is applicable. The values of $\delta$ found with Kogut-Susskind quarks are much smaller than 0.12 but consistently positive. The Kogut-Susskind fits are done at smaller $m_\pi/m_\rho$ than those with Wilson fermions, and therefore would correspond to smaller values of $m_q$ if measured in common units. Thus there is some evidence for a slow approach to the behavior of Eq. (4), perhaps with $\delta$ of 0.12, for $m_\pi/m_\rho$ sufficiently below 0.3. In any case, if the extrapolations to small quark mass to be discussed in Sect. 4.1 were done using Eq. (3) for $m_\pi/m_\rho$ above 0.3 and using Eq. (4) with $\delta$ of 0.12 for $m_\pi/m_\rho$ below 0.3, in place of the actual fits using Eq. (3) everywhere, the changes in predicted

| lattice | $\beta$ | $m_\pi/m_\rho$ | $\delta$ |
|---|---|---|---|
| Wilson | | | |
| $16^3 \times 24$ | 5.70 | $0.49 - 0.69$ | $0.04(6)$ |
| $24^3 \times 36$ | 5.93 | $0.47 - 0.74$ | $-0.03(3)$ |
| $30 \times 32^2 \times 40$ | 6.17 | $0.48 - 0.74$ | $-0.07(3)$ |
| Kogut-Susskind | | | |
| $32^3 \times 64$ | 6.00 | $0.31 - 0.50$ | $\approx 0.03$ |
| $32^3 \times 64$ | 6.50 | $0.40 - 0.66$ | $0.02(1)$ |

Table 3
Values of the parameter $\delta$ obtained from fits to data for the pion mass as a function of quark mass.

hadron mass ratios would be smaller than the statistical errors.

Labrenz and Sharpe [19] have also attempted to use chiral perturbation theory to estimate the effect of the valence approximation on the nucleon mass, $m_N$. While I believe that the basic idea of trying to use chiral perturbation theory to make such comparisons is very promising, the particular results which they presented in Dallas appear to me to be unconvincing. For the behavior of $m_N$ at small pion mass, chiral perturbation theory for full QCD gives an expression of the form

$$m_N = a + b_\pi m_\pi^2 + b_K m_K^2 + c_\pi m_\pi^3 + c_K m_K^3. \tag{6}$$

Chiral perturbation theory for the valence approximation gives an expression with an additional pion term and no $K$ terms, since these entail vacuum strange quark loops,

$$m_N = a' + a''m_\pi + b'_\pi m_\pi^2 + c'_\pi m_\pi^3. \tag{7}$$

The extra $a''m_\pi$ term arises from the same mechanism leading to Eq. (4), and the coefficient $a''$ is proportional to $\delta$. Our earlier discussion of $\delta$ suggests that the $a''$ coefficient will be small.

Although it might seem that a comparison of Eqs. (6) and (7) could be used to estimate the error in valence approximation calculations of $m_N$, it turns out this can not be done. One difficulty is that Eq. (6) describes the variation of $m_N$ as both sea and valence quark masses are varied together, while Eq. (7) describes only the result of varying the valence quark mass with the sea quark mass kept fixed. As the sea quark mass varies in full QCD, the effective low-momentum gauge coupling varies, and therefore the gauge coupling to be put into the corresponding valence approximation must vary. Thus for each choice of $m_\pi$ and $m_K$ in the full QCD relation Eq. (6), a different and a priori unknown value of $a'$ must be used in the corresponding valence approximation relation Eq. (7). For lattice spacing sufficiently small, a change in the gauge coupling entering the valence approximation causes a change in low lying hadron masses by a single overall scale factor. Thus the ratio of Eq. (7) to a corresponding equation for some other mass parameter, such as $f_\pi$, will be free of the scale factor and perhaps might be compared to the ratio of Eqs. (6) to an expression for $f_\pi$ in full QCD.

A useful comparison for the ratio $m_N/f_\pi$ still can not be made. The difficulty at this point is that the valence approximation chiral perturbation theory expression for $m_N/f_\pi$ depends on an unknown additive constant, comparable to $a'$ in Eq. (7), giving the valence approximation to $m_N/f_\pi$ in the limit of zero quark mass. I know of no way to fix this number from chiral perturbation theory. In the limit of very heavy $m_\pi$ and $m_K$, $m_N/f_\pi$ will become equal in full QCD and in the valence approximation. This limit, however, is beyond the range of applicability of chiral perturbation theory and provides no help in determining the missing parameter. One might try to determine the missing coefficient by a fit to the data of Ref. [6]. Unfortunately, as mentioned already in the discussion of Eq. (4), this data does not appear to extend to sufficiently small quark mass for the full apparatus of chiral perturbation theory to be applicable.

If, nonetheless, the expression of Eq. (7) is fitted [19] to the data of Ref. [6] at $\beta$ of 5.93, the extrapolated value of $m_N$ at the light quark mass differs from the result of the linear extrapolation used in Ref. [6] by only about 3%.

There are comparisons between full QCD and the valence approximation which have fewer difficulties. An example of one such comparison is for $f_K/f_\pi$. In both full QCD and in the valence approximation this quantity must become 1 if $m_\pi$ is made equal to $m_K$. The required additive constants are thus determined. A completely reliable comparison between $f_K/f_\pi$ in full QCD and in the valence approximation depends on unknown higher terms in the chiral lagrangians in both cases. With plausible guesses for these terms, Sharpe [20] obtains the result that this quantity in the valence approximation will be smaller than in full QCD by about 0.12. The calculation of Ref. [24] finds $f_K/f_\pi$ below its experimentally observed value by 0.06(9).

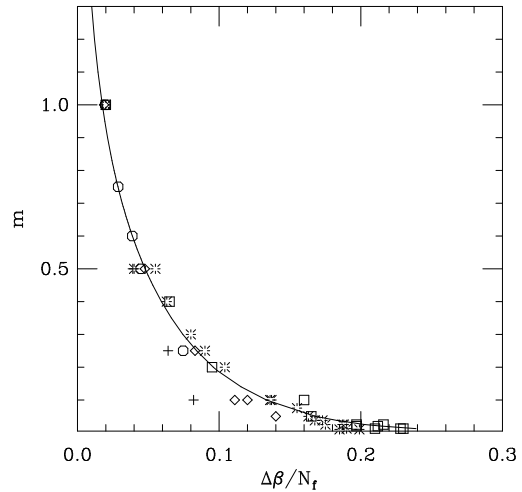

Figure 4. Monte Carlo data for the shift in beta of the deconfining transition from the valence approximation to full QCD, for Kogut-Susskind fermions, in comparison to the one loop weak-coupling prediction.

Still another method of calculating differences between the valence approximation and full QCD has been considered by Hasenfratz and De-Grand [21]. These authors evaluate the effective shift in chromoelectric charge arising from vacuum polarization due to heavy Kogut-Susskind quark-antiquark pairs in full QCD. For sufficiently heavy quarks, the determination of this change is reduced to evaluating the one loop quark-antiquark vacuum polarization term coupling two gluons in the weak-coupling expansion for lattice QCD. The resulting shift in $\beta$ is linear in the number of quark flavors. The predicted $\Delta\beta$ is compared with Monte Carlo data for the difference between the critical $\beta_c$ of the deconfining transition for QCD without quarks and $\beta_c$ with a number of quarks ranging from 2 to 24. For lattices with time direction period ranging from 4 to 8, Figure 4 shows the predicted shift in comparison to Monte Carlo data. The overall agreement is quite good for $m_q a \geq 0.05$.

## 2.4. Wave Functions, State Vectors

I would like to mention briefly two recent pieces of work on hadron wave functions and state vectors. Kieu and Negele [22] have examined a Bethe-Salpeter wave function for the pion with quark and antiquark joined, in effect, by the ground state chromoelectric field configuration which would occur between a static color charge at the quark position and static anticolor charge at the antiquark position. They found that this wave function falls off with the quark-antiquark separation significantly more slowly than does the Coulomb gauge wave function. Thus for increasing quark-antiquark separation, a quark-antiquark pair joined by the ground state field configuration becomes progressively closer to the true pion state vector than does a Coulomb gauge quark-antiquark pair.

Liu [23] has compared hadron masses calculated in the usual valence approximation with masses obtained in the valence approximation using a quark coupling matrix that allows quarks to propagate in only one direction of time. The state vector of a hadron in the usual valence approximation, Liu shows, at any instant may include quark-antiquark pairs in addition to the valence quarks. These pairs do not occur in hadrons composed of quarks which can propagate in only one time direction. As consequence of eliminating these pairs, Liu finds, the rho and pion become nearly degenerate and the nucleon and delta baryon become nearly degenerate.

## 3. Spectrum Calculations with Fixed Lattice Spacing

## 3.1. Valence Approximation

Three recent spectrum calculations using the valence approximation are summarized in Table 4. Kim and Sinclair used Kogut-Susskind fermions with wall sources and point sinks. The UKQCD collaboration used the clover action. The hadron propagators were found with point sources and point sinks. The clover action [29] consists of the Wilson fermion action with the simplest additional term added to cancel, to zeroth order in the chromoelectric coupling constant, the O(a) irrelevant coupling of the Wilson action.

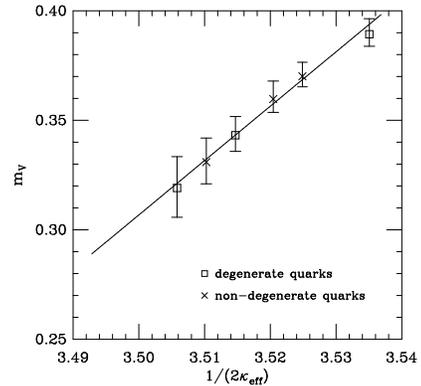

Figure 5. UKQCD data for vector mesons masses. For quark and antiquark hopping constants $k_1$ and $k_2$, $k_{eff}^{-1}$ is $(2k_1)^{-1} + (2k_2)^{-1}$.

In addition to calculating pseudoscalar and vector masses with quark and antiquark masses equal, the UKQCD collaboration also found masses for mesons composed of a quark and antiquark with different masses. With this data a direct test can be made of Eq. (2) and the corresponding equation for the squares of pseudoscalar meson masses. Figure 5 shows a comparison of the masses of vector mesons composed of quark and antiquark with the same mass and the masses of vector mesons composed of quark and antiquark with different masses. The data strongly supports Eq. (2). Corresponding data for pseudoscalar masses supports the version of Eq. (2) for squared pseudoscalar masses.

| | action | lattice | $\beta$ | $m_q a$ | $m_\pi/m_\rho$ | configs. | ref. |
|---|---|---|---|---|---|---|---|
| Argonne | valence, KS | $32^3 \times 64$ | 6.0 | 0.0100 | 0.500(7) | 66 | [25] |
| | | $32^3 \times 64$ | 6.0 | 0.0050 | 0.393(10) | 66 | [25] |
| | | $32^3 \times 64$ | 6.0 | 0.0025 | 0.311(9) | 66 | [25] |
| | | $32^3 \times 64$ | 6.5 | 0.0100 | 0.659(8) | 100 | [25] |
| | | $32^3 \times 64$ | 6.5 | 0.0050 | 0.520(12) | 100 | [25] |
| | | $32^3 \times 64$ | 6.5 | 0.0025 | 0.401(15) | 100 | [25] |
| | action | lattice | $\beta$ | k | $m_\pi/m_\rho$ | configs. | ref. |
| UKQCD | valence, clover | $24^3 \times 48$ | 6.2 | 0.14144 | 0.77(1) | 60 | [26] |
| | | $24^3 \times 48$ | 6.2 | 0.14226 | 0.62(2) | 60 | [26] |
| | | $24^3 \times 48$ | 6.2 | 0.14262 | 0.52(2) | 60 | [26] |
| GF11 | valence, Wilson | $16^3 \times 32$ | 5.70 | 0.1600 | 0.856(3) | 219 | [6] |
| | | | | 0.1650 | 0.690(6) | 219 | [6] |
| | | | | 0.16625 | 0.612(6) | 219 | [6] |
| | | | | 0.1675 | 0.491(8) | 219 | [6] |
| | | $24^3 \times 32$ | 5.70 | 0.1600 | 0.854(2) | 92 | [6] |
| | | | | 0.1650 | 0.693(4) | 92 | [6] |
| | | | | 0.1663 | 0.610(5) | 58 | [6] |
| | | | | 0.1675 | 0.502(6) | 92 | [6] |
| | | $24^3 \times 36$ | 5.93 | 0.1543 | 0.830(5) | 210 | [6] |
| | | | | 0.1560 | 0.737(6) | 210 | [6] |
| | | | | 0.1573 | 0.603(6) | 210 | [6] |
| | | | | 0.1581 | 0.466(8) | 210 | [6] |
| | | $32^2 \times 30 \times 40$ | 6.17 | 0.1500 | 0.867(2) | 219 | [6] |
| | | | | 0.1519 | 0.735(4) | 219 | [6] |
| | | | | 0.1526 | 0.633(6) | 219 | [6] |
| | | | | 0.1532 | 0.478(1) | 219 | [6] |
| | action | lattice | $\beta$ | k | $m_\pi/m_\rho$ | time | ref. |
| HEMCGC | full QCD, Wilson | $16^3 \times 32$ | 5.3 | 0.1670 | 0.722(4) | 2400 | [27] |
| | | $16^3 \times 32$ | 5.3 | 0.1675 | 0.599(8) | 1300 | [27] |

Table 4
Parameters of recent hadron spectrum calculations.

## 3.2. Full QCD

Data for full QCD with two flavors of Wilson fermions collected by the HEMCGC collaboration [27] is also listed in Table 4. These calculations use the hybrid Monte Carlo algorithm. Hadron propagators were calculated for wall sources and either wall or point sinks, and in addition for Gaussian sources with either Gaussian or point sinks. The integrated autocorrelation time for the plaquette was found to be about 80 time units for the data at the heavier pion mass and 120 time units for the data at the lighter pion mass. An examination of the dispersion of pion effective masses calculated from propagators averaged over bins gave roughly equivalent autocorrelation estimates. Thus the data collected at the larger pion mass consisted of perhaps 30 independent configurations and the data with the smaller pion mass perhaps 10. Figures 6 and 7 show full QCD results obtained with various combinations of sea quark and valence quark hopping constants in comparison to valence approximation masses for Wilson fermions on a lattice $16^3 \times 32$ at $\beta$ of 5.85 and 5.95 [28]. For the range of parameters considered, the differences between the valence approximation and full QCD consistent with the statistical uncertainties.

## 4. Valence Approximation Zero Lattice Spacing Limits

### 4.1. Masses

The first systematic evaluation of the infinite volume continuum limit of hadron masses extrapolated to physical quark masses was reported this year by the GF11 collaboration [6]. Hadron masses were calculated with Wilson quarks on a set of four lattices. The parameters entering these calculations are listed in Table 4.

Each configuration was fixed into Coulomb gauge. In all cases, hadron propagators were then calculated using a Gaussian quark source with mean radius squared $<r^2>$ of 6 in lattice units, and using point sinks and Gaussian sinks with $<r^2>$ of 1.5, 6, 13.5, and 24. Hadron masses were fitted to each channel by examining effective mass plots and selecting the largest interval which might appear to be a plateau. An automatic fit-ting program then did a fit to each subwindow of four or more time slices within this region. The subwindow giving the smallest value of $\chi^2$ per degree of freedom was selected as the final fit. Fits to each window were done by minimizing the fit's correlated $\chi^2$ among all data points fitted. Fits were also done in this way simultaneously to the propagators for point sinks and Gaussian sinks with $< r^2 >$ of 1.5 and 6. Error bars on all of these results were found by the bootstrap method.

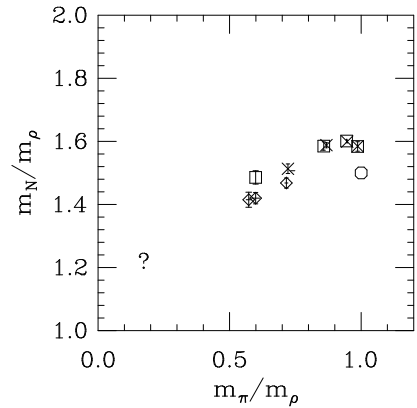

Figure 6. HEMCGC mass ratios with various combinations of valence quark and sea quark hopping constants. The squares have sea quark $k$ of 0.1675, the crosses have sea quark $k$ of 0.1670. The diamonds are valence approximation results. The circle and question mark show expected values for infinite quark mass and for the real world value of the light quark masses.

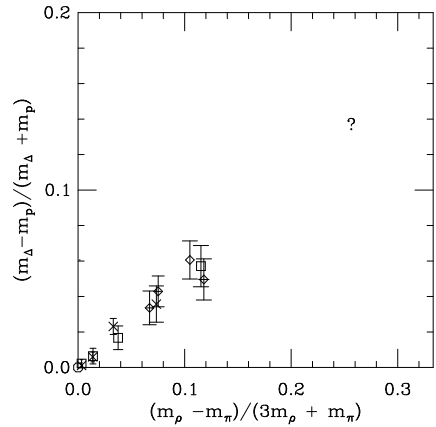

Figure 7. HEMCGC hyperfine splittings. Symbols have same meaning as in preceding figure.

A examination of effective mass plots showed, in all cases, that the propagators for hadron sinks with larger $< r^2 >$ reached plateaus at smaller time separations between source and sink and therefore coupled more weakly to excited states. Thus a comparison of masses fitted to channels with different radii provided a test for the presence of high mass contamination. Typical effective mass plots and fits are shown in Figures 8, 9 and 10. These figures show the rho on $30 \times 32^2 \times 40$ with k of 0.1532, for sinks with $< r^2 >$ of 0, 6 and 24. Table 5 gives the masses obtained from these fits, along with masses from fits to sinks with $< r^2 >$ of 1.5 and 13.5, and the mass from a simultaneous fit to data with $< r^2 >$ of 0 (point sinks), 1.5 and 6. The consistency of these results is evidence that the lowest mass has been picked out accurately in each channel. Similar results are obtained from propagators found on the other GF11 lattices listed in Table 4. Thus the masses determined from these fits appear to be free of the high mass contamination which Ukawa [5] suggested might occur.

For each GF11 data set of Table 4, the critical hopping constant $k_c$ was determined, as usual, by fitting $m_\pi^2$ in lattice units to $C[(2k)^{-1} - (2k_c)^{-1}]$. With quark mass $m_q$ then defined as $(2k)^{-1} - (2k_c)^{-1}$, hadron masses for a continuous range of quark masses were found by fitting the Monte Carlo data to linear functions of quark mass. Figure 11 shows such fits to hadron masses at the three lightest values of $m_q$ for the lattice $30 \times 32^2 \times 40$ at $\beta$ of 6.17. The masses were determined by simultaneous fits to propagators with sink $< r^2 >$ of 0, 1.5 and 6. The up and down quark masses were then taken to be equal and given by the "normal" quark mass, $m_n$, which was determined by fitting the ratio of extrapo-lated values, $m_\pi/m_\rho$ to experiment. The strange quark mass was found using the pseudoscalar version of Eq. (2) to determine $m_K$, then fitting the ratio of extrapolated values $m_K/m_\rho$ to experiment. Masses for other strange hadrons were found using Eq. (2) and its analogues for baryons.

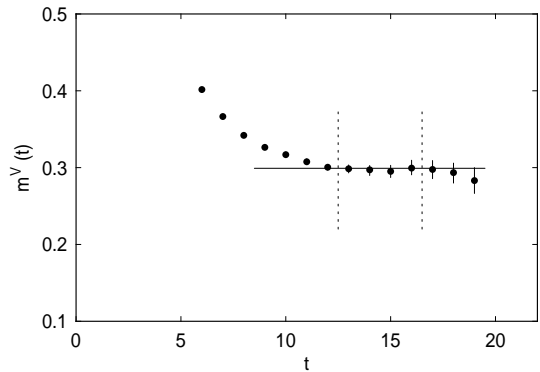

Figure 8. Effective masses and fitted mass for the rho propagator with point sink on the lattice $30 \times 32^2 \times 40$ with k of 0.1532. Dotted vertical lines bound the window selected by the automated fitting procedure.

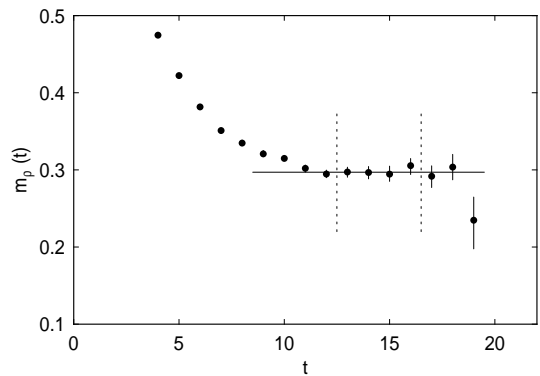

Figure 9. Effective masses and fitted mass for the rho propagator with Gaussian sink $< r^2 >$ of 6. on the lattice $30 \times 32^2 \times 40$ with k of 0.1532. Dotted vertical lines bound the window selected by the automated fitting procedure.

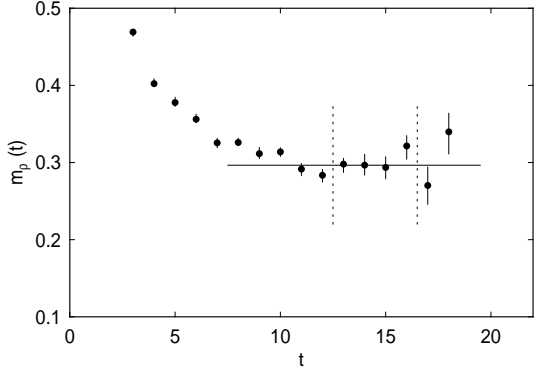

Figure 10. Effective masses and fitted mass for the rho propagator with Gaussian sink $< r^2 >$ of 24. on the lattice $30 \times 32^2 \times 40$ with k of 0.1532. Dotted vertical lines bound the window selected by the automated fitting procedure.

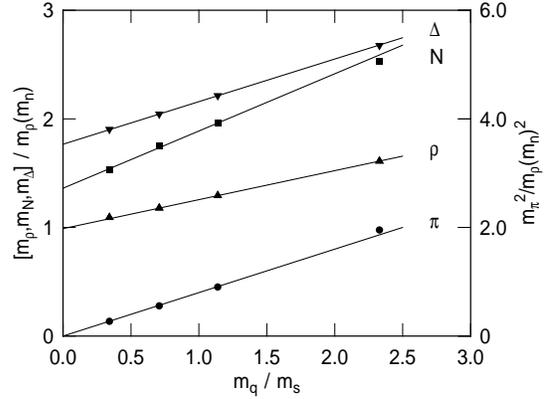

Figure 11. For a $30 \times 32^2 \times 40$ lattice at $\beta$ of 6.17, $m_\pi^2$, $m_\rho$, $m_N$ and $m_\Delta$, in units of the physical rho mass $m_\rho(m_n)$, as functions of the quark mass $m_q$, in units of the strange quark mass $m_s$. The symbol at each point is larger than the error bars.

| hadron | $< r^2 >$ | mass | range | $\chi^2$ |
|---|---|---|---|---|
| pion | 0 | 0.1445(20) | 12-15 | 0.2 |
| | 1.5 | 0.1453(20) | 15-18 | 0.4 |
| | 6 | 0.1451(19) | 13-17 | 0.1 |
| | 13.5 | 0.1453(20) | 13-16 | 0.0 |
| | 24 | 0.1455(20) | 13-16 | 0.0 |
| rho | 0,1.5,6 | 0.3024(48) | 13-16 | 0.4 |
| | 0 | 0.2991(59) | 13-16 | 0.2 |
| | 1.5 | 0.2987(59) | 12-15 | 0.1 |
| | 6 | 0.2969(66) | 13-16 | 0.1 |
| | 13.5 | 0.2965(76) | 13-16 | 0.0 |
| | 24 | 0.2965(88) | 13-16 | 0.0 |
| nucleon | 0,1.5,6 | 0.4234(65) | 13-16 | 1.1 |
| | 0 | 0.4187(59) | 12-15 | 0.5 |
| | 1.5 | 0.4188(87) | 12-17 | 0.7 |
| | 6 | 0.4199(88) | 13-16 | 0.5 |
| | 13.5 | 0.4215(80) | 13-16 | 0.2 |
| | 24 | 0.4097(78) | 10-13 | 0.9 |
| delta | 0,1.5,6 | 0.5267(81) | 13-17 | 0.9 |
| | 0 | 0.5233(88) | 14-17 | 0.1 |
| | 1.5 | 0.5188(111) | 14-17 | 0.0 |
| | 6 | 0.5133(104) | 14-17 | 0.5 |
| | 13.5 | 0.5197(101) | 11-16 | 0.9 |
| | 24 | 0.5120(101) | 11-15 | 0.7 |

Table 5
Fitted masses from the lattice $32 \times 30 \times 32 \times 40$ with k of 0.1532

For Wilson fermions, hadron mass ratios are expected to approach their continuum limits as linear functions of lattice spacing. Linear extrapolations of hadron mass ratios to zero lattice spacing are shown in Figures 12 - 14. The masses in this extrapolations were found from simultaneous fits to sinks with $< r^2 >$ of 0, 1.5 and 6 for the runs on $16^3 \times 32$ with $\beta$ of 5.7, $24^3 \times 36$ with $\beta$ of 5.93 and $30 \times 32^2 \times 40$ with beta of 6.17. The values of $\beta$ in each case are chosen so that the lattice period $L$ is the same in physical units in all cases and within statistical errors equal to $9m_\rho^{-1}$.

The continuum ratios found in finite volume were then corrected to infinite volume. After extrapolation to physical quark mass, hadron masses measured in units of $m_\rho$ on the lattice $16^3 \times 32$ were compared with ratios on the lattice $24^3 \times 32$. The changes were all found to be less than 4.5%. Using either Lüscher's formula [11] for the approach of masses to their infinite volume limit or the model of Ref. [10], it can then

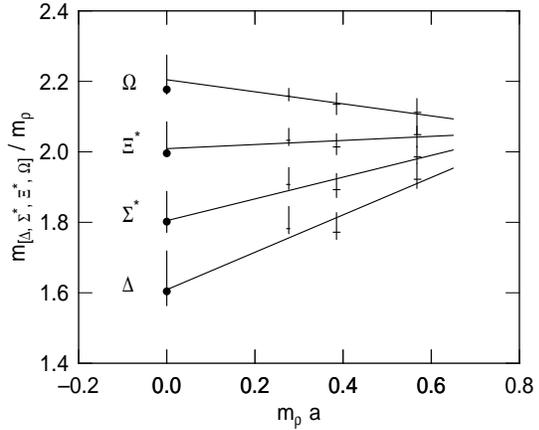

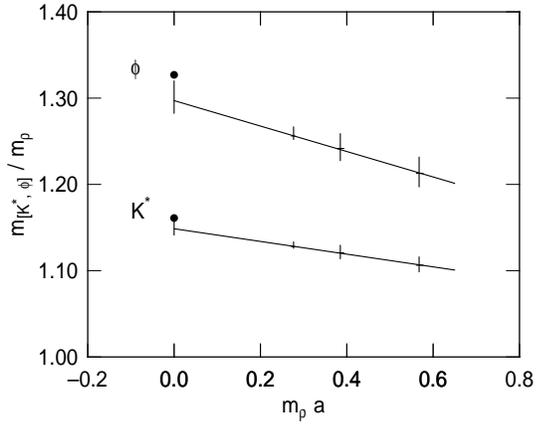

Figure 12. Mass ratios and linear fits as functions of the lattice spacing $a$ in units of $1/m_\rho$. The error bars at zero lattice spacing are uncertainties in the extrapolated ratios, and the points at zero lattice spacing are observed values.

Figure 14. Mass ratios and linear fits as functions of the lattice spacing $a$ in units of $1/m_\rho$. The error bars at zero lattice spacing are uncertainties in the extrapolated ratios, and the points at zero lattice spacing are observed values.

be argued that masses on $24^3$ are within 1% of their infinite volume limit.

For any hadron mass $m$ measured in units of $m_\rho$ define the finite volume correction term for lattice period $L$

$$\Delta(\beta, L) = \frac{m}{m_\rho}(\beta, \infty) - \frac{m}{m_\rho}(\beta, L). \quad (8)$$

The quantity needed is $\Delta(\infty, 9m_\rho^{-1})$. Based on the size of the change in $m/m_\rho$ from $\beta$ of 5.7 to infinity and on the size of the change in this ratio from $24^3 \times 32$ to infinite volume, it is then argued

$$\Delta(\infty, 9m_\rho^{-1}) \approx \frac{m}{m_\rho}(5.7, 9m_\rho^{-1}) - \frac{m}{m_\rho}(5.7, 13.5m_\rho^{-1}), \quad (9)$$

to within an error of less than 2% of $m/m_\rho$. Infinite volume continuum mass ratios were then found using Eq. (9).

Table 6 lists continuum hadron mass ratios found in finite volume and after applying the infinite volume correction. The infinite volume numbers taken together are statistically consistent with experiment. The differences between these numbers and experiment are less than 6%,

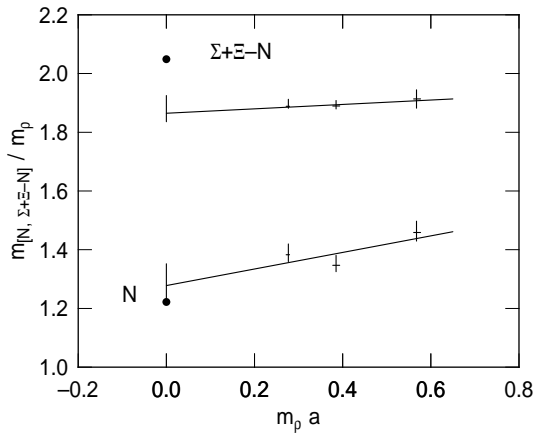

Figure 13. Mass ratios and linear fits as functions of the lattice spacing $a$ in units of $1/m_\rho$. The error bars at zero lattice spacing are uncertainties in the extrapolated ratios, and the points at zero lattice spacing are observed values.

| ratio | fin. vol. | inf. vol. | obs. |
|---|---|---|---|
| $m_{K^*}/m_\rho$ | 1.149(9) | 1.166(16) | 1.164 |
| $m_\Phi/m_\rho$ | 1.297(19) | 1.333(32) | 1.327 |
| $m_N/m_\rho$ | 1.278(68) | 1.216(104) | 1.222 |
| $\Delta m/m_\rho$ | 1.865(45) | 1.927(74) | 2.047 |
| $m_\Delta/m_\rho$ | 1.609(78) | 1.565(122) | 1.604 |
| $m_{\Sigma^*}/m_\rho$ | 1.805(58) | 1.806(80) | 1.803 |
| $m_{\Xi^*}/m_\rho$ | 2.009(49) | 2.055(65) | 1.996 |
| $m_\Omega/m_\rho$ | 2.205(56) | 2.296(89) | 2.177 |
| $\Lambda_{\overline{MS}}^{(0)}/m_\rho$ | 0.305(8) | 0.319(12) | |

Table 6
Calculated hadron mass ratios in comparison to observed. The mass difference $\Delta m$ is $m_\Xi + m_\Sigma - m_N$.

| decay | fin. vol. | inf. vol. | obs. |
|---|---|---|---|
| $f_\pi/m_\rho$ | 0.106(9) | 0.106(14) | 0.121 |
| $f_K/m_\rho$ | 0.121(6) | 0.123(9) | 0.148 |
| $F_\rho/m_\rho$ | 0.177(21) | 0.173(29) | 0.199 |
| $F_\phi/m_\rho$ | 0.217(19) | 0.253(35) | 0.219 |

Table 7
Calculated values of meson decay constants extrapolated to zero lattice spacing in finite volume, then corrected to infinite volume, compared with observed values.

equivalent to less than 1.6 times the statistical error.

The value of $\Lambda_{\overline{MS}}^{(0)}/m_\rho$ shown is obtained by extrapolation to zero lattice spacing of $(\Lambda_{\overline{MS}}^{(0)}a)/(m_\rho a)$ as a linear function of $m_\rho a$. Here $\Lambda_{\overline{MS}}^{(0)}a$ is determined from $\alpha_{\overline{MS}}$ by the two-loop Callan-Symanzik equation, and $\alpha_{\overline{MS}}$ is found from $\beta$ following Ref. [30]. The slope of the fit of $(\Lambda_{\overline{MS}}^{(0)}a)/(m_\rho a)$ to a linear function of $m_\rho a$ is statistically consistent with zero. Thus $m_\rho a$ follows asymptotic scaling within statistical errors.

## 4.2. Decay Constants

The first evaluation of the infinite volume continuum limits of meson decay constants was reported by the GF11 group in Ref. [24]. The calculation used the same the data sets listed in Table 4. The fits to propagators and extrapolations to physical quark mass, zero lattice spacing and infinite volume were all done by versions of the methods used in the mass evaluation of Ref. [6]. Table 7 shows predicted results in comparison to experiment. The finite renormalizations factors in these numbers are found to one-loop order following the mean-field improved perturbation expansion of Lepage and Mackenzie [31]. Statistically equivalent results but with slightly larger error bars are given by zero order mean-field improved finite renormalizations. Equivalent results with still larger error bars are given by naive finite renormalizations.

Although overall the predicted values are in fair agreement with experiment, the circumstance that three of four decay constants range from 0.9 to 2.8 standard deviations below experiment, suggests that valence approximation decay constants may lie systematically somewhat below experiment. A simple physical argument supports this possibility [30]. The valence approximation can be thought of as the replacement of the momentum and frequency dependent color dielectric constant caused by quark-antiquark vacuum polarization with its low momentum limit [13]. At low momentum the effective quark charge in the valence approximation agrees with the low momentum effective charge of the full theory. The valence approximation could therefore be fairly reliable for low-lying baryon and meson masses if, as seems likely, these masses are determined largely by the long distance behavior of the chromoelectric field. Due to the absence of dynamical quark-antiquark vacuum polarization, however, the quark charge in the valence approximation falls faster with momentum than it does in the full theory. As a result at short distance the attractive quark-antiquark potential in the valence approximation is weaker than in the full theory, and meson wave functions in the valence approximation are smaller at the origin less than in the full theory. Meson decay constants are proportional to the square of wave functions at the origin and thus should be to be smaller in the valence approximation than in the full theory.

## 4.3. Glueball Masses

An evaluation of the infinite volume, continuum limit of glueball masses was reported by a Liverpool-Wuppertal collaboration [32]. The lattice sizes, ensemble sizes and $\beta$ of the new calcu-lation are shown in Table 8. This calculation used the same methods as earlier work by Michael and Teper at smaller values of $\beta$ [33]. Glueball operators were constructed by an iterative gauge invariant smearing process which, at each step, returns smeared links which have twice the length of the links taken as input [34]. For each time separation between source and sink operators, a matrix of correlation functions was found among the various operators carrying the same quantum numbers. The possible operators for each set of quantum numbers were made by summing together rotations of loops with the same shape. The maximal eigenvalue of this matrix was then extracted. The logarithm of the ratio of maximal eigenvalue at separation $t$ to the maximal eigenvalue at separation $t+1$ defines the effective mass at separation $t$. By including a range of different shapes of loops in the set of operators, then maximizing the eigenvalue of the correlation matrix, combined operators were produced which appear to couple very efficiently to the lightest state in each channel. An comparison of effective masses produced this way and with simpler choices of glueball operator suggests that simpler choices do not select out the lightest states as efficiently.

| lattice | $\beta$ | configs. | $m_{0^{++}}a$ |
|---|---|---|---|
| Liverpool-Wuppertal | | | |
| $16^3 \times 32$ | 6.0 | 3040 | 0.701(27) |
| $32^4$ | 6.2 | 1001 | 0.522(38) |
| $32^4$ | 6.4 | 3200 | 0.415(14) |
| GF11 | | | |
| $16^3 \times 24$ | 5.70 | 8094 | 0.983(40) |
| $20^3 \times 30$ | 5.83 | 4002 | 0.858(43) |
| $16^3 \times 24$ | 5.93 | 30640 | 0.786(12) |
| $12^3 \times 24$ | 5.93 | 48278 | 0.771(10) |
| $24^3 \times 36$ | 6.17 | 31150 | 0.582(10) |
| $30 \times 32^2 \times 40$ | 6.40 | 25440 | 0.433(11) |

Table 8
Parameters of recent glueball spectrum calculations.

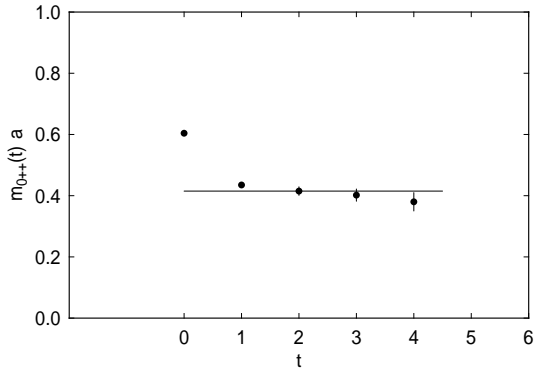

Figure 15. Effective masses and final mass value found by the Liverpool-Wuppertal group for the $0^{++}$ glueball on a lattice $32^4$ at $\beta$ of 6.4.

The final mass value in each channel was found generally by looking for the smallest time separation giving an effective mass statistically consistent with the effective masses at larger time separation. Figure 15 shows effective masses and the mass taken as the final value for the $0^{++}$ glueball at $\beta$ of 6.4. In this case, the effective mass between $t$ of 2 and 3 was chosen as the final mass value. Although this choice appears reasonable, it is not clear whether the data has actually reached a plateau at this point. It appears possible that the rate of growth in error bars may be obscuring a continued decline in the effective mass. A large collection of masses were also calculated for other glueball states. The effective mass plots again did not show clear plateaus. The final masses in most other channels were therefore taken only as upper bounds.

To evaluate the continuum limit, $m_{0^{++}}a$ in lattice units was divided by the square root of the string tension in lattice units $\sqrt{K}a$. This ratio is expected to approach its continuum limit linearly in $a^2$. A linear fit was made to the ratios $(m_{0^{++}}a)/(\sqrt{K}a)$ as a function of $Ka^2$. The conversion of mass ratios to MeV was done using an estimate of $K$ in physical units of 440 MeV, based on a variety of indirect arguments. The value of $m_{0^{++}}$ at $\beta$ of 6.4 was found to be about 50 MeV below the continuum limit of 1600 MeV. The mass at 6.4, $1550 \pm 50$ MeV, was then quoted as the continuum limit with a additional systematic extrapolation error of 50 MeV. No estimate was given for the error associated with the choice of value of $K$.

An alternate way to obtain a continuum limit prediction for $m_{0^{++}}$ is to fit the ratio

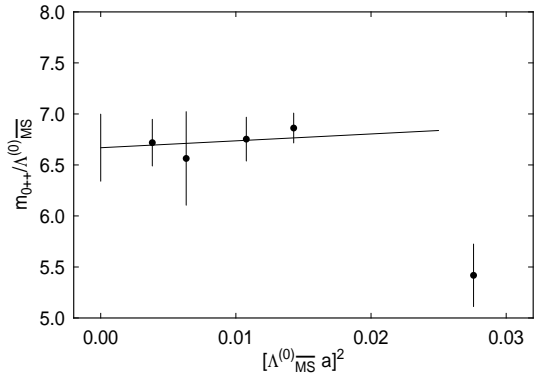

Figure 16. Liverpool-Wuppertal data for $(m_{0^{++}}a)/(\Lambda^{(0)}_{\overline{MS}}a)$ and a linear extrapolation to zero lattice spacing.

$(m_{0^{++}}a)/(\Lambda^{(0)}_{\overline{MS}}a)$ to a linear function of $(\Lambda^{(0)}_{\overline{MS}}a)^2$. A value of $\Lambda^{(0)}_{\overline{MS}}$ in physical units is given by the continuum limit of $\Lambda^{(0)}_{\overline{MS}}/m_\rho$ reported in Ref. [6]. This value is free of the ambiguities in the estimation of the physical value of $K$. Figure 16 shows the continuum limit of $m_{0^{++}}$ obtained in this way. Data at lower $\beta$ is also included from Refs. [33] and [35]. The predicted continuum $m_{0^{++}}/\Lambda^{(0)}_{\overline{MS}}$ is $6.67\pm0.33$. The value of $\Lambda^{(0)}_{\overline{MS}}$ from the continuum limit of $(\Lambda^{(0)}_{\overline{MS}}a)/(m_\phi a)$ is $243.7\pm6.8$ MeV. This is statistically consistent with the value found using $m_\rho$ but has a somewhat smaller error bar. Including the uncertainty in $\Lambda^{(0)}_{\overline{MS}}$, the final prediction obtained from the Liverpool-Wuppertal data for $m_{0^{++}}$ is then $1625\pm92$ MeV [36].

A calculation of the infinite volume continuum limit of glueball masses was also reported by the GF11 collaboration [37]. The lattice sizes and $\beta$ used are shown in Table 8. For the lattice $20^3\times30$ results are shown for smeared glueball operators defined in Coulomb gauge. For the four other lattices gauge invariant smeared glueball operators are constructed following Ref. [38]. Smeared links are defined by adding to each link the sum of space direction staples with a coefficient $\epsilon$ which was taken to be 0.25 at $\beta$ of 5.7 and 1.0 otherwise. This step is then iterated some number of times $N_S$. From the smeared links, square loops $N_L\times N_L$ were constructed. Sums over rotations of such loops then give operators for the $0^{++}$, $1^{+-}$ and $2^{++}$ glueball. A correlation matrix among the operators with each set of quantum numbers for a range of $N_S$ and $N_L$ was evaluated at each time separation. Effective masses were determined from the diagonal terms in these matrices and intervals of smearing sizes and loop sizes found for which effective mass plots showed clear plateaus. Figure 17 shows the effective mass plot for the $0^{++}$ glueball at $\beta$ of 6.4 found for $N_S$ of 8 and $N_L$ of 9. A clearer plateau occurs here than in Figure 15 since the ensemble used for Figure 17 is about eight times larger. Similar plateaus in $0^{++}$ effective masses were found for several other combinations of $N_S$ and $N_L$ for this value of $\beta$ and for the other values listed in Table 8. A mass was extracted from the propagator of Figure 17 fitting the time interval indicated. Masses were found in other channels similarly. An optimal mass was then found among the set of channels showing plateaus using a correlation matrix among these masses generated by the bootstrap method. Alternative fits using the full underlying propagator matrix gave statistically equivalent results.

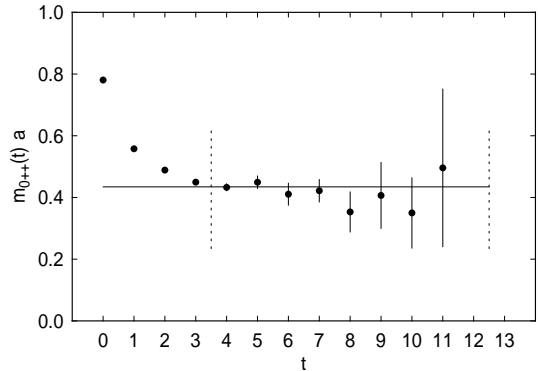

Figure 17. Effective masses and final mass value found by the GF11 group for the $0^{++}$ glueball on a lattice $30\times32^2\times40$ at $\beta$ of 6.4.

For the $2^{++}$ the effective mass plateaus were weaker than those for the $0^{++}$ and below $\beta$ of 5.93 a clear signal could not be found. For the $1^{+-}$ glueball an acceptable signal could not be found in any of the data sets of Table 8.

Figure 18 shows the masses found for

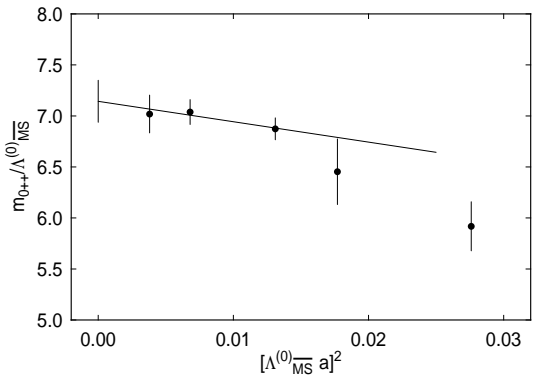

Figure 18. GF11 date for $(m_{0^{++}}a)/(\Lambda^{(0)}_{\overline{MS}}a)$ and a linear extrapolation to zero lattice spacing.

the $0^{++}$ glueball, and an extrapolation of $(m_{0^{++}}a)/(\Lambda^{(0)}_{\overline{MS}}a)$ to zero lattice spacing. The predicted continuum $m_{0^{++}}/\Lambda^{(0)}_{\overline{MS}}$ is $7.14 \pm 0.21$. Using $\Lambda^{(0)}_{\overline{MS}}$ from the continuum limit of $(\Lambda^{(0)}_{\overline{MS}}a)/(m_\phi a)$ and including the uncertainty in $\Lambda^{(0)}_{\overline{MS}}$, the final predicted $m_{0^{++}}$ is $1740 \pm 71$ MeV.

A comparison of the GF11 data for the lattice $16^3 \times 24$ and $12^3 \times 24$ at $\beta$ of 5.93 combined with Lüscher's formula for volume dependence [11] gives a bound of 0.5% on the difference between this prediction and its infinite volume limit. A similar estimate applies to the Liverpool-Wuppertal value of $m_{0^{++}}$.

The Liverpool-Wuppertal values for $m_{0^{++}}$ at fixed lattice spacing, and for the continuum limit both in units of $\Lambda^{(0)}_{\overline{MS}}$ and in MeV are statistically consistent with the GF11 values. The GF11 values, however, appear to be more reliable because of the larger ensembles used and the resulting more clearly defined effective mass plateaus. The two well established $0^{++}$ states within any reasonable distance of $1740 \pm 71$ MeV are $f_0(1590)$ and $f_0(1710)$. The first of these is 2 standard deviations low thus $f_0(1710)$ appears to be favored.

A question which remains is why a valence approximation glueball calculation should be believed at all. A commonly held opinion is that the $0^{++}$ may mix strongly with quark-antiquark states and might be shifted by as much as 20% in going to full QCD. My guess, on the other hand, is that the mass shift due to mixing is likely to be less than 4%. A $0^{++}$ quark-antiquark state must have odd orbital angular momentum. The resulting angular momentum barier should tend to suppress the quark-antiquark annihilation required for mixing with chromoelectric field. An estimate of the effect on $m_{0^{++}}$ of mixing with quark-antiquark states is given by the mass shift in quark-antiquark states, with nonzero orbital angular momentum, due to mixing with chromo-electric field. In the valence approximation the isosinglet $2^{++}$ state $f_2(1270)$ and isovector $2^{++}$ state $a_2(1320)$ are exactly degenerate. In the real world, the $f_2$ mixes with glue and the $a_2$ does not. Thus the splitting between the $f_2$ and $a_2$ is the shift in the $f_2$ due to mixing. This splitting, and therefore the shift, is $43 \pm 5$ MeV. Similar estimates can be made comparing $h_1(1170)$ with $b_1(1235)$, $f_1(1285)$ with $a_1(1260)$, $\omega_3(1670)$ with $\rho_3(1690)$, and perhaps also comparing $f_0(975)$ with $a_0(980)$ if these are quark-antiquark states rather than kaon molecules. The splittings between these pairs range from $9 \pm 3$ MeV to $60 \pm 20$ MeV. Thus it seems to me reasonable to guess that the shift in the mass of the $0^{++}$ glueball due to mixing with quark-antiquark states will be less than about 60 MeV, which is 3.4%.

The other possible source of error in the valence approximation to the $0^{++}$ mass arises from virtual quark-antiquark loops contributing more than just a shift in the effective chromoelectric charge. This same error, however, is present in the valence approximation to the masses of hadrons composed of quarks and antiquarks. The effect of this error on low-lying hadron masses, discussed in Sect. 4.1, is less than 6%. This error combined with the effect of mixing could decrease $1740 \pm 71$ MeV enough to make it consistent with $f_0(1590)$. The $f_0(1710)$ would still appear to be favored.

I am grateful to a large number of people for answering my email, explaining their work, and sending data and pictures. I would particularly like to thank G. Bali, T. DeGrand, S. Gottlieb, Y. Iwasaki, T. Kieu, Y. Kim, J. Labrenz, D. Leinweber, K.F. Liu, J. Negele, F. Rapuano, D. Richards, K. Schilling, S. Sharpe, A. Ukawa, A.

Vaccarino, and T. Yoshie.